\documentstyle[preprint,prd,aps,12pt]{revtex}

\preprint{LAVAL-PHY-98-11}
\tighten

\begin{document}
\author{B. Dion, L. Marleau and G. Simon}
\address{D\'epartement de Physique, Universit\'e Laval\\
Qu\'ebec QC Canada, G1K 7P4}
\title{Scalar and Vector Leptoquark Pair Production\\
at Hadron Colliders:\\
Signal and Backgrounds}
\date{1998}
\maketitle
\draft

\begin{abstract}
We perform a systematic analysis of scalar and vector leptoquark pair
production at the Fermilab Tevatron and at the CERN LHC. We evaluate signal
expectations and background levels for the processes $pp (p \bar{p})
\rightarrow$ 2 jets$\; + \;e^{+}e^{-}$ and 2 jets$\;+\;e + {\not{p}}_T$. The
Monte Carlo event generator ISAJET is used to simulate the experimental
conditions at the current ($\sqrt{s}=1.8$ TeV, ${\cal L}=100\;$pb$^{-1}$)
and upgraded (${\cal L}=2\;$fb$^{-1}$) Tevatron as well as the LHC ($\sqrt{s}%
=14$ TeV, ${\cal L}=10\;$fb$^{-1}$). Depending on the luminosity, and
assuming a branching ratio $B(LQ\rightarrow eq)=0.5$, we find a discovery
reach up to 170 (255) GeV for scalar leptoquarks at the current (upgraded)
Tevatron. Similarly, we find vector leptoquarks to be detectable at masses
below $300$ ($400$) GeV depending on the coupling. At the LHC, the discovery
reach is enhanced to $1 \;$TeV for scalar leptoquarks and to $1.5 \;$TeV for
vectors.
\end{abstract}


\section{Introduction}

Leptoquarks are color-triplet particles carrying both baryon and lepton
quantum numbers which mediate transitions between the quark and lepton
sectors. They appear in many extensions of the Standard Model (SM) such as
composite models\cite{comp} which postulate a common preonic substructure to
quarks and leptons, Grand Unified Theories\cite{gut,ps} which treat quarks
and leptons on the same basis, superstring-inspired E$_{6}$ models\cite
{string,hr} and strongly-coupled Abbott-Fahri models\cite{af}. They arise as
scalar ($S=0$) as well as vector ($S=1$) or even fermionic ($S=1/2$)
particles depending on the model considered. Leptoquarks could be produced
in $e^{+}e^{-}$ \cite{ee,opal1,opal2,L3}, $e\gamma $ \cite{eg}, $ep$ \cite
{ep,H1,ZEUS} and hadron colliders \cite{cdf,d0,norman,bg,MMS,sigvec}.

Recently, the H1\cite{H1} and ZEUS\cite{ZEUS} experiments at HERA have
reported an excess of large $Q^{2}$ deep inelastic scattering events
compared to QCD expectations. One possible explanation which has been put
forward to account for these events is that they arise from the single
production of a $\sim 200\;$GeV leptoquark. The statistics for these high-$%
Q^{2}$ events remain quite low for now (H1 finds 8 events leading to a
leptoquark mass $M\sim 200$ GeV while ZEUS finds 4 events with $M\sim 220$
GeV) and the latest results from both collaborations\cite{hera98} seem to
indicate no significant deviations from Standard Model expectations in the
1997-1998 run. It nonetheless seems worthwhile to investigate the discovery
potential of both scalar and vector leptoquarks of such mass at the Tevatron
and at the CERN LHC especially since their production in pairs at hadron
colliders is almost insensitive to an unknown Yukawa coupling.

In addition, leptoquark searches are also going on at the Fermilab Tevatron
and at LEP. For instance, the OPAL collaboration\cite{opal1} at LEP recently
fixed a limit of $M_{LQ}\geq $ 131 GeV for scalar leptoquarks of the first
generation. Similarly, the OPAL\cite{opal2} and L3\cite{L3} collaborations
have investigated the indirect production of scalar and vector leptoquark in
the $t$-channel and as contact interactions in the $e^{+}e^{-}\rightarrow q%
\bar{q}$ channel. Their results are expressed in terms of the ratio $%
M_{LQ}/\lambda $ where $\lambda $ is the Yukawa coupling of the leptoquark
to the quark-lepton pair. They find a limit $\lambda /e\leq 0.2-0.7$ for
leptoquarks of mass $\sim $ 200 GeV. On the other hand, the latest results
from the CDF \cite{cdf} and D0 \cite{d0} collaborations at the Tevatron
exclude scalar leptoquarks with masses below 225 GeV and 204 GeV for
branching ratios of the leptoquarks to the electron equal to 1 and 0.5
respectively. Making use of their published search for the superpartner of
the top quark, the D0 collaboration has also looked for leptoquarks which
decay exclusively into neutrinos. Their analysis yields a limit of 79 GeV in
this case.

Previous studies have suggested that present Tevatron data \cite{d0} most
likely excludes vector leptoquarks with masses below 250 GeV. However, a
comprehensive study of the various Standard Model backgrounds and the extent
to which the signal of a 200-250 GeV leptoquark will be reduced by the
kinematic cuts imposed on these backgrounds was still lacking up to now.
This motivated an important part of the work presented in this paper, i.e.
to present such an analysis and evaluate the vector leptoquark discovery
reach at the Tevatron and the LHC. For that purpose, we implement vector
leptoquark data and related cross sections in the ISAJET\cite{isajet} event
generator and evaluate the importance of the LQ signal and the SM
backgrounds. The ISZRUN package contained in the ZEBRA version of ISAJET is
used to perform the selection cuts.

Earlier analyses on pair production of leptoquarks at hadron are available.
Only some aspect or conditions have been examined. For example, the authors
have analyzed the pair production of scalar leptoquarks at hadron colliders 
\cite{dms1,dms2} focussing on the cleanest of the possible signatures, i.e.
2 jets $+e^{+}e^{-}$. Similarly, other contributions examined pair and
single production of scalar leptoquarks at the LHC\cite{eboli1} and single
production of scalar and vector leptoquarks through effective photon
approximations at the LHC\cite{eboli2}. In this paper, we investigate the
ability of the Fermilab Tevatron and at the CERN LHC to observe the pair
production of both scalar and vector leptoquarks. We perform a comprehensive
analysis which includes the evaluation of the cross section as well as the
optimization of the signal-to background ratio through the use of selection
and kinematic cuts. We assume a branching ratio $B(LQ\rightarrow eq)=0.5$
and consider the following two signatures for leptoquark pair production: 2
jets $+e^{+}e^{-}$ or 2 jets $+e+{\not{p}}_{T}.$ Finally, we estimate the
discovery reach of scalar and vector leptoquarks at the Tevatron and at the
LHC.

\section{Event simulation}

The Monte-Carlo event generator ISAJET is used to model the experimental
conditions at the Tevatron and at the LHC. The toy detector used for the
Tevatron is loosely based on the D0 and CDF detectors at Fermilab. The
calorimeter is segmented in cells of size $\bigtriangleup \eta \times
\bigtriangleup \phi =0.1\times 0.0875$ with $\eta $ coverage $-4<\eta <4$,
where $\eta $ is the rapidity and $\phi $ is the azimuthal angle. The
hadronic energy resolution is defined such as the jet resolution is $70\%/%
\sqrt{E}$ while the electromagnetic energy resolution is $15\%/\sqrt{E}$.

The experimental conditions at the LHC are also simulated, with the ATLAS%
\cite{atlas} and CMS\cite{cms} detectors in mind. The calorimeter is divided
in cells of dimension $\bigtriangleup \eta \times \bigtriangleup \phi
=0.05\times 0.05$, with a rapidity range of $-5<\eta <5$. The hadronic
energy resolution is given by

\begin{itemize}
\item  $50\%/\sqrt{E}\oplus 0.03$ for $-3<\eta <3$,

\item  $100\%/\sqrt{E}\oplus 0.07$ for $3<\mid \eta \mid <5$.
\end{itemize}

The electromagnetic energy resolution is given by $10\%/\sqrt{E}\oplus 0.01$
independently of the rapidity.

Jets are found using a fixed cone algorithm with radius $R=\sqrt{%
(\bigtriangleup \eta )^{2}+(\bigtriangleup \phi )^{2}}=0.7$. The jet energy
scale is set by requiring

\begin{itemize}
\item  a transverse energy $E_{T}>15$ (50) GeV at the Tevatron (LHC)

\item  a pseudorapidity $|\eta _{j}|\leq 1.5$.
\end{itemize}

On the other hand, electrons are required to

\begin{itemize}
\item  be separated from any jet by $\Delta R\geq 0.3$

\item  have a transverse momentum $p_{T}>20$ (100) GeV at the Tevatron (LHC)

\item  have a pseudorapidity $|\eta _{l}|\leq 1.5$.
\end{itemize}

The calculations are performed using the CTEQ3M parton parametrization\cite
{PDFLIB}. Higher-order corrections to the cross section are available for
pair of scalar leptoquarks \cite{Zerwas}. The calculations presented here
always assume the lowest-order cross sections. Higher-order corrections
could be applied in the form of a $K$-factor whenever available.

\section{Signal and backgrounds}

\subsection{Leptoquark signal}

For the purposes of this work, we do not focus on any specific model and
consider the production of generic first-generation leptoquarks which can be
of type $(eu)$ (with electric charge $Q=$$\pm \frac{1}{3}$, $\pm \frac{5}{3}$%
) or $(ed)$ (with $Q=$$\pm \frac{2}{3}$, $\pm \frac{4}{3}$) which decay into
an electron and a quark or into a neutrino and a quark with equal branching
ratios $B(LQ\rightarrow eq)=B(LQ\rightarrow \nu q)=0.5$ or in an electron
and a quark with $B(LQ\rightarrow eq)=1$. After hadronization and
fragmentation, the expected signatures for leptoquark pair production
consist of two (or more) jets along with two electrons, one electron and a
missing transverse momentum or simply a missing transverse momentum.

Leptoquark pair production occurs via two dominant subprocesses, $q\bar{q}$
annihilation or gluon fusion (see Fig.1 for the Feynman diagrams). The only
model-dependent parameter is the Yukawa coupling which appears in the $t$%
-channel diagram in $q\bar{q}$ annihilation. Low-energy data as well as
collider experiments constrain this coupling to be of the order of electric
strength or lower. In this case, the contribution coming from this diagram
becomes negligeable with respect to the others. Consequently, the
corresponding contribution is not included in the cross section. This
absence of sensivity to the magnitude of the Yukawa coupling, along with the
large center-of-mass energy, is one of the main reasons why it is worthwhile
to investigate pair production of leptoquarks in hadron colliders.

The cross sections for both scalar and vector leptoquark pair production in
hadron colliders have been extensively reviewed\cite{bg,MMS,sigvec}. In the
scalar case, the leptoquark coupling to the gluon is simply given by the
strong coupling. There is however an ambiguity in the vector case. In order
to evaluate the $q\bar{q}$, $gg\rightarrow VV$ cross sections both the
trilinear $gVV$ and the quartic $ggVV$ couplings need to be evaluated. In
any model in which vector leptoquarks appear as fundamental objects, they
correspond to the gauge bosons of an extended gauge group. In this case both
couplings are fixed by gauge invariance. However, there is a possibility
that vector leptoquarks arise as a low-energy manifestation of a more
fundamental theory at larger scale, in which case some anomalous couplings
may appear. One such coupling comes from the ``anomalous magnetic moment'',
usually described by the parameter $\kappa $ and which takes unit value in
the gauge theory case. The Lagrangian takes the form:\newline
\begin{equation}
{\cal L}_{V}=-{\frac{1}{2}}F_{\mu \nu }^{\dagger }F^{\mu \nu
}+M_{V}^{2}V_{\mu }^{\dagger }V^{\mu }-ig_{s}\kappa V_{\mu }^{\dagger
}G^{\mu \nu }V_{\nu } \\
\end{equation}
where $G_{\mu \nu }$ is the usual gluon field strength tensor, $V_{\mu }$ is
the leptoquark field and $F_{\mu \nu }=D_{\mu }V_{\nu }-D_{\nu }V_{\mu }$.
One might also include an anomalous electric quadrupole moment \cite{sigvec}
usually described by the variable $\lambda $ and which takes the value $%
\lambda =0$ in the gauge case. Here, for the sake of clarity, we present our
analysis for only two specific cases: $\kappa =\lambda =0$ (minimal
coupling) and $\kappa =1$, $\lambda =0$ (gauge coupling). The cross section
for vector leptoquark pair production at hadron colliders has previously
been calculated for various sets of parameters $\kappa $ and $\lambda $ \cite
{sigvec}. For $\lambda =0$, the results show that the cross section is
largest in the gauge case ($\kappa =1$) and reaches a minimum for values of $%
\kappa \sim 0$. Both differ by a factor somewhat less than an order of
magnitude. By comparison, the cross section for pair production of scalar
leptoquark turns out to be smaller by a factor of $\sim 2-3$ compared with
vector leptoquarks with minimal coupling.

\subsection{SM Backgrounds}

The main sources of background to leptoquark pair production as identified
by \cite{bg,MMS} are (1) gauge boson production involving the production of
two jets, (2) QCD processes involving the production of heavy flavors ($b$, $%
c$) and (3) $t\bar{t}$ production.

\subsubsection{Gauge boson production}

The main background to leptoquark pair production at the Tevatron comes from 
$Z^{*}+ 2$ jets (Fig.2(a)) in the dielectron channel and from $W^{*}+ 2$
jets (Fig.2(b)) in the electron plus missing transverse energy channel\cite
{cdf,d0,dms1,dms2,eboli1}.In the latter case, the source of hard electrons
comes from the leptonic decay of the gauge bosons while the jets typically
arise from gluon radiation from incoming partons. Although potentially
large, this background can be reduced by an invariant mass cut on the lepton
pair (dielectron case) or a transverse mass cut on the lepton and missing $%
p_T$ (single electron case). The present analysis will show that this
background can also be reduced by kinematic cuts on the transverse energy of
the jets and leptons.

\subsubsection{QCD processes}

Background from QCD processes (Fig.2(c)) arise from the hadroproduction of
heavy quarks $c$ and $b$. Note that the top quark is not included in this
list. The wide difference in mass results in a completely distinct signature
and topology and the top quark will be dealt with separately. Relevant QCD
processes include pair production ($c \bar c$, $b \bar b$), single
production ($c q $, $c g$, $b q $, $b g$) and gluon pair production with one
of the gluons leading to $c \bar c$ or $b \bar b$. This background can in
principle be important as the relative cross sections are very large.
However, a great majority of QCD events can be eliminated by adequate cuts.
In particular, QCD events typically have much lower transverse momentum
distributions than $t \bar t$ and leptoquark production events. This is
particularly true in the case of the leptons which in this case arise from
the semileptonic decay of the $c$ and $b$ hadrons.

\subsubsection{$t\bar{t}$ background}

Pair production of top quarks (Fig.2(d)) will certainly be a dominant source
of background at the LHC. This background implies the semileptonic decay of
one of the quarks into a $b$ quark, an electron and a neutrino, $\nu _{e}$.
This process is characterized by a missing transverse momentum, which
implies that the electron energy is lower than that of the corresponding
jet. Another potential background comes from the single production of a top
quark accompanied by a quark or a gluon ($t\,q(g)\rightarrow t\,q(g)$).
However, previous results have shown\cite{eboli1} this background to be
negligeable with respect to the pair production channel. The value $M_{t}=$
175 GeV is assumed for the top mass.

\section{Selection cuts}

\label{sigback} Leptoquark pair production events are generated assuming a
branching ratio $B(LQ\rightarrow eq)=0.5$ unless stated otherwise and
classified according to the number of jets and of isolated leptons present
in each event. As discussed above, the signature to leptoquark pair
production can consist of

\begin{enumerate}
\begin{enumerate}
\item  2 jets + $e^{+}e^{-}$ (dielectron channel),

\item  2 jets + $e+{\not{p}}_{T}$ (single electron channel),

\item  2 jets + ${\not{p}}_{T}$ (missing $p_{T}$ channel).
\end{enumerate}
\end{enumerate}

Typically, channels (a) and (c) occur 25\% of the time while channel (b)
occurs with a 50\% probability. The jets and leptons are expected to be
emitted in the central rapidity region with high transverse energy.
Leptoquark events originating from signal (c) tend to be overwhelmed by the
QCD background coming from $q\bar{q}$ and gluon pair production and do not
allow for a clean identification as the lepton-quark invariant mass cannot
be reliably calculated in this case. Consequently we restrict ourselves to
signals (a) and (b). This means that $\sim 75$\% of all the leptoquark pair
production events can contribute to the final results while the other 25\%
are lost.

The selection cut on the events corresponding to channel (a) is made by
requiring

\begin{enumerate}
\item  two isolated electrons with a transverse energy $E_{T}>20$ (100) GeV
and a pseudorapidity $|\eta _{e}|\leq 1.5$ at the Tevatron (LHC),

\item  at least two jets with a transverse energy $E_{T}>15$ (50) GeV and a
pseudorapidity $|\eta _{j}|\leq 1.5$ at the Tevatron (LHC).
\end{enumerate}

Events corresponding to channel (b) are required to contain

\begin{enumerate}
\item  one isolated electron with a transverse energy $E_{T}>25$ (100) GeV
and a pseudorapidity $|\eta _{e}|\leq 1.5$ at the Tevatron (LHC),

\item  at least two jets with a transverse energy $E_{T}>25$ (50) GeV and a
pseudorapidity $|\eta _{j}|\leq 1.5$ at the Tevatron (LHC).
\end{enumerate}

An optimization is carried out by varying kinematic cuts until they maximize
the signal-to-backgrond ratio. The nature and values of those cuts are based
on earlier calculations \cite{cdf,d0,norman,dms1,dms2}. For instance, in
order to reduce the background coming from Drell-Yan processes we impose a
cut on the invariant mass of the lepton pair. We thus eliminate all events
near the $Z$ peak: $82$ GeV $\leq M_{e^{+}e^{-}}\leq 100$ GeV. We do not
apply a transverse mass cut on the lepton and missing $p_{T}$ to reduce the
background from $W$ events in the (b) channel but choose instead to require
a missing transverse energy ${\not{p}}_{T}\geq 40$ (200) GeV at the Tevatron
(LHC).

Finally, we impose a cut on the total transverse energy of the jets and
leptons as defined by 
\begin{equation}
S_{T}=\sum_{j}E_{T}^{j}+\sum_{e}E_{T}^{e}.
\end{equation}
Note that the former expression is based on the convention used in \cite
{norman} and should not be confused with the $S_T$ variable used in other D0
publications which includes the missing transverse momentum. We find the
optimal cuts at the Tevatron to be $S_{T}\geq $200 and 190 GeV for events
corresponding to channel (a) and (b) respectively. Similarly we find $%
S_{T}\geq $1.5 and 1.0 TeV to be optimal at the LHC.

\section{Leptoquark production at the Tevatron}

The results for the Tevatron are shown in Figs 3-6. Figure 3 displays the
missing $p_{T}$ distributions in the single electron channel for the various
backgrounds as well as for vector leptoquarks in the gauge case ($\kappa =1$%
). The selection cuts described in Section \ref{sigback} have been applied
and the value $B(LQ\rightarrow eq)=0.5$ has been chosen for the leptoquark
branching ratio into an electron and a quark. The corresponding figures for
the minimal coupling and scalar cases are not displayed as they are very
similar, differing roughly by a constant factor. One sees from Fig. 3 that
the background for QCD processes is dominant at small values of ${\not{p}}%
_{T}$, as could be presumed from the large cross section as well as the
typically low transverse momentum available for the neutrino in this case.
On the other hand, the background from $W+$ jets and $Z+$ jets becomes
important at slightly higher values of ${\not{p}}_{T}$, around 50 -100 GeV.
The background from $t\bar{t}$ is comparatively negligeable and is
practically invisible in Fig.3. The ${\not{p}}_{T}$ distribution of the
leptoquark signal shows an approximately constant behavior for values of ${%
\not{p}}_{T}\leq 150$ GeV and gradually decreases for higher values. In
order to reduce the QCD background sufficiently without significantly
reducing the signal, we impose a cut of 40 GeV on ${\not{p}}_{T}$.

Figure 4 shows the $S_{T}$ distribution for the signal as well as the
various backgrounds once the ${\not{p}}_{T}$ cut has been applied. Both
single electron and dielectron channels have been included in this graph.
Once again, only the results for the gauge case are displayed. Clearly, the $%
W+$ jets, $Z+$ jets and QCD backgrounds, though very important, are peaked
at small values of $S_{T}$ ($\sim $100 GeV for $W, Z \, +$ jets and $\sim $%
150 GeV for QCD) while the leptoquark peak is wider and concentrated at
higher values of $S_{T}$. Roughly, one can see that the peak corresponds
typically to twice the leptoquark mass, especially for leptoquark masses
approaching the kinematic limit of the collider. This is to be expected as
the lepton-quark transverse energy then corresponds approximately to the
leptoquark mass. An interesting feature is that the peak widens as the
leptoquark mass increases. This feature is largely due to the increasing jet
multiplicity at higher values of the leptoquark mass. For instance, about
40\% of the events corresponding to $M_{LQ}\geq 250$ GeV contain 3 or 4 jets
while this proportion is about 25\% for masses below 200 GeV. These excess
jets, which are emitted in the fragmentation/hadronization phase of the
event, escape with a fraction of the available transverse energy and cause
the observed smearing. We find the overall optimal cut to be of 190 GeV for
the single electron channel and 200 GeV for the dielectron channel.

Figure 5 (a-c) displays the invariant mass distribution of both signal and
background for ${\not{p}}_{T}>200$ GeV and $S_{T}>190$ (200) GeV for the
single electron (dielectron) channel. The lepton-jet pairing algorithm is
based on the transverse energy of the quarks and leptons. Specifically, the
largest-$p_{T}$ electron is paired with the lowest-$E_{T}$ jet and
vice-versa. For single-electron events, the same approach is applied with
the ${\not{p}}_{T}$ used as corresponding to the neutrino transverse energy.
This figure shows the strong quark-lepton correlation in the leptoquark
signal. The mass peak is quite narrow and allows for a straightforward
identification of the leptoquark mass.

One can see from Fig. 5 that the background is dominated by $W, Z\, +$ jets
processes which come mostly from the production of a $W^{\pm }$ in the
single-electron channel and from $Z^{0}$ in the dielectron channel. Note
that transverse mass cuts can also be applied to reduce the background from $%
W\, +$ jets \cite{d0}. However, the application of this cut seems
unnecessary here as the background is most important only at small values of
the quark-lepton invariant mass ($\sim 75-150$ GeV). On the other hand,
backgrounds from QCD and $t\bar{t}$ production are relatively small and are
also concentrated at small values of the invariant mass. Leptoquarks of
masses 200-250 GeV and higher are thus virtually background-free.

The leptoquark discovery reach at the Tevatron in the scalar as well as in
the vector cases can be read directly from Fig. 6. In this figure, the
partial cross section for leptoquark pair production is displayed. This
graph is obtained by integrating over the lepton-jet invariant mass peak for
the whole range of leptoquark masses. For leptoquarks of mass lower than 200
GeV, we integrate over a bin of width $\Delta M_{ej}=40$ GeV around the peak
while we take $\Delta M_{ej}=50$ GeV for $M_{LQ}\geq 200$ GeV.

In this figure, both the $B=0.5$ (Fig. 6 (a)) and $B=1$ (Fig. 6 (b)) cases
are included for the leptoquark branching ratio. A closer look at these
graphs reveals that the partial cross section is larger when the branching
ratio is equal to unity. This is partly due to the fact that all the
leptoquark events observed in this case correspond to the dielectron channel
and can thus be included in the final results, in contrast with the $B=0.5$
case where 25\% of the events correspond to the ${\not{p}}_{T}$ channel
(channel (c) of Section \ref{sigback}) and are excluded from the analysis. A
second explanation comes from the relative efficiency of the selection cuts
to single out events in both channels. Indeed, a thorough analysis of our
results indicates that the ratio of events that are selected from the sample
tends to be larger in the dielectron channel than in the single electron
channel.

The backgrounds are also illustrated in Fig. 6. While the dominant
background comes from $W,Z\,+$ jets production and QCD processes in the $%
B=0.5$ case, the only remaining background when $B=1$ arises from Drell-Yan
production of a $Z\,+$ jets. In both cases, this background is small in
comparison with the signal and is concentrated at small values of the
electron-jet invariant mass. We impose a $5\sigma $ statistical significance
as well as a minimum of 5 events for leptoquark discovery. The integrated
luminosity for the Tevatron current run is 100 pb$^{-1}$ and the
corresponding limit on the cross section is illustrated by the dashed line
of Fig. 6. We find a discovery reach of 170 (175) GeV for scalar leptoquarks
with a branching ratio of $B=\;$0.5 (1). Comparison with present Tevatron
data can be achieved by relaxing the requirements to a $5\sigma $
statistical significance relative to the background. Considering the current
accumulated luminosities of 123 and 115 pb$^{-1}$ in the dielectron and
single electron channels at D0, we find a reach of 205 (215) GeV for $B=\;$%
0.5 (1) in agreement with the results of D0. In the case of vector
leptoquarks, the discovery reach for 5 events is enhanced to 225 (235) GeV
with the minimal coupling $\kappa =0$ and 280 (290) GeV with the Yang-Mills
coupling $\kappa =1$ for $B=\;$0.5 (1).

It is also interesting to reestimate these bounds in view of the luminosity
upgrade which is supposed to take place when the Main Injector comes into
function in 2000-2002. The expected luminosity should reach 2 fb$^{-1}$\cite
{tev33}. Processes with cross sections above the dotted line of Fig. 6 could
then be seen. This allows for a discovery reach of 255 (265) GeV for scalar
leptoquarks with a branching ratio of $B=$0.5 (1). Similarly, we find vector
leptoquarks to be detectable up to 315 (325) GeV for a minimal anomalous
coupling and 365 (380) GeV for a gauge-like coupling.

\section{Leptoquark production at the LHC}

The results of our analysis for the LHC are displayed in Figs 7-12. Let us
first examine the ${\not{p}}_{T}$ distribution for backgrounds and vector
leptoquark signal in Fig. 7. As above, $B=0.5$ is used for the branching
ratio of the leptoquark into an electron and a quark and only the single
electron channel is included in the graph. The main backgrounds can be seen
to arise from QCD processes and $t\bar{t}$ production. Backgrounds from $W$
and $Z$ production are eliminated by the selection cuts of Sect. \ref
{sigback}. Typically, the electrons emitted in these processes have rather
low transverse momentum as can be expected from the fact that the $W$ and $Z$
bosons are generally collinear with the beam. Clearly, this distribution
shows that the neutrino emitted in semileptonic decay of $c$ and $b$ hadrons
does have a small transverse energy. This background can thus be
significantly reduced by applying a ${\not{p}}_{T}$ cut. The $t\bar{t}$
distribution displayed in the same figure is peaked at higher values and
becomes dominant over the QCD background at large ${\not{p}}_{T}$. In
contrast, the neutrinos which originate from leptoquark decay typically
possess a much larger transverse energy independently of the leptoquark
mass. This is explained by the fact that the neutrino radiated by the $W$
boson in the decay of the top quark shares the available transverse energy
with an electron while the whole transverse energy is transmitted to the
neutrino in leptoquark decay. On the other hand, the ${\not{p}}_{T}$
distribution of the leptoquark signal is approximately constant up to $\sim
1 $ TeV for $M_{LQ}=750$ GeV and higher for $M_{LQ}\geq $ 1 TeV. A ${\not{p}}%
_{T}$ cut of 200 GeV is adequate to optimize the signal-to-background ratio.

Figure 8 shows the $S_{T}$ distribution for QCD and $t\bar{t}$ background as
well as for gauge-like vector leptoquarks of masses 750, 1000 and 1250 GeV.
One sees from this figure that the main backgrounds originate from QCD
events and from $t\bar{t}$ production. The leptoquark signal is clearly
dominant independently of the $S_{T}$ cut. This is in contrast to the
Tevatron case where this cut was essential to discriminate the signal from
the (mostly electroweak) background. This domination of the leptoquark
signal is caused by the ${\not{p}}_{T}$ cut of 200 GeV we have applied
beforehand. Finally, one sees once again that the $S_{T}$ distribution for
the background is peaked at significantly smaller values than the signal.
The peak is located in both background cases at values around 1 TeV while
the signal is characterized by a wider distribution peaked at values over 2
TeV. Clearly, a $S_{T}$ cut of $\sim 2$ TeV would be sufficient to eliminate
the background completely. However, such a cut would also reduce the
signal-to-background ratio as the leptoquark signal takes relatively large
values in part of this range. Moreover, the background is already quite
small. We choose instead the more moderate cuts of $S_{T}>1$ (1.5) TeV in
the single electron (dielectron) channels.

Figures 9 - 11 show the lepton-jet invariant mass distributions where the ${%
\not{p}}_{T}$ and $S_{T}$ cuts have been applied. Again, a branching ratio $%
B=0.5$ is assumed and the contributions from the dilelectron and single
electron channels are added. The same transverse-energy algorithm as above
is used to pair electron and jets. One notes that the QCD and $t\bar{t}$
backgrounds have practically vanished thanks to the ${\not{p}}_{T}$ and $%
S_{T}$ cuts, leaving the leptoquark signal virtually background-free. With
this in mind, the $5\sigma $ statistical significance we require is
trivially satisfied. The only requirement left is on the number of events,
which is once again taken to be five.

While the background is seen to be negligeable in the low-mass regime, the $t%
\bar{t}$ background reappears at higher values of the lepton-jet invariant
mass, see Fig 9 (c-d) and Figs. 10-11 (e-f). However, the absence of
correlation in the invariant mass $M_{ej}$ for this background combined with
a clean and rather narrow peak in the leptoquark signal allows the $5\sigma $
requirement to be satisfied.

One comment is in order regarding the invariant distributions of Figs. 9 -
11 (a). We see that the leptoquark mass peak is accompanied by a smeared
bump at higher values of the invariant mass for low-mass leptoquark. This
feature can be explained by the use of the transverse energy as criteria for
lepton-jet pairing. This method works better when the kinematic limit for
leptoquark production is approached. Leptoquarks of lower mass are generally
emitted with much higher transverse momentum and the corresponding
lepton-jet pair is thus emitted with high $p_{T}$ and small angular
separation. The uncertainty on the transverse momentum is thus increased
while the signature will consist of one electron-jet pair in one hemisphere
and another pair in the opposite hemisphere. The ideal pairing algorithm in
this case would be to pair each electron with the nearest-neighbor jet
instead. Nevertheless, as the total cross section for leptoquark is very
large in the low-mass region, this alternative is not necessary.

Let us now consider the leptoquark discovery reach at the LHC. Once again,
Fig. 12 shows the partial cross section for leptoquark production. As above,
we have integrated the lepton-jet invariant mass distribution around the
leptoquark peak for the whole kinematic range of leptoquark masses. The bin
width over which we integrate varies as the peak widens and we take

\begin{itemize}
\item  $M_{LQ}\leq 500$ GeV: $\Delta M_{ej}=100$ GeV

\item  $500$ GeV$<M_{LQ}\leq 1$ TeV: $\Delta M_{ej}=150$ GeV

\item  $1$ TeV$<M_{LQ}\leq 1.5$ TeV: $\Delta M_{ej}=200$ GeV

\item  $M_{LQ}>1.5$ TeV: $\Delta M_{ej}=250$ GeV.
\end{itemize}

The results are displayed in Fig. 12 (a-b) for the scalar and vector cases
with branching ratios of $B=0.5$ (1). The LHC luminosity is assumed equal to
10 fb$^{-1}$. The corresponding discovery limit for 5 events is illustrated
by the dashed line in Fig. 12. We find a discovery reach for scalar
leptoquarks of 1 (1.1) TeV with $B=0.5$ (1). In the vector case, the
corresponding reach is 1.3 (1.4) TeV for $\kappa =0$ and 1.55 (1.65) TeV for 
$\kappa =1$.

\section{Summary}

Summarizing, we have presented the results of a complete analysis of
first-generation scalar and vector leptoquark pair production at the
Tevatron and the LHC. Both the dielectron and the single electron channels
are included in the simulations. The importance of the various Standard
Model backgrounds with the same signature is also evaluated. We found the
gauge boson background ($W, Z \, +$ 2 jets) to be dominant at the Tevatron,
followed by the background from QCD processes. For the LHC however, the
background is relatively small and dominated by QCD processes and $t\bar{t}$
production. The discovery limits are established by requiring a $5\sigma $
statistical significance as well as a minimum of 5 events in a one-year run.
Considering an integrated luminosity of ${\cal L}=100$ pb$^{-1}$ (${\cal L}%
=2 $ fb$^{-1}$) at the Tevatron, the results show that scalar leptoquarks
can be detected up to 175 (265) GeV depending on the branching ratio.
Similar limits on vector leptoquarks are set to 235 (325) GeV in the minimal
anomalous coupling case ($\kappa =0$) and 290 (380) GeV in the gauge case ($%
\kappa =1$). The LHC discovery potential for leptoquarks increases these
limits to 1 TeV in the scalar case, and to 1.3 and 1.55 TeV for vectors with
anomalous magnetic couplings $\kappa =0$ and 1 respectively.

\acknowledgements
This research was supported by the Natural Sciences and Engineering Research
Council of Canada and by the Fonds pour la Formation de Chercheurs et l'Aide
\`{a} la Recherche du Qu\'{e}bec.

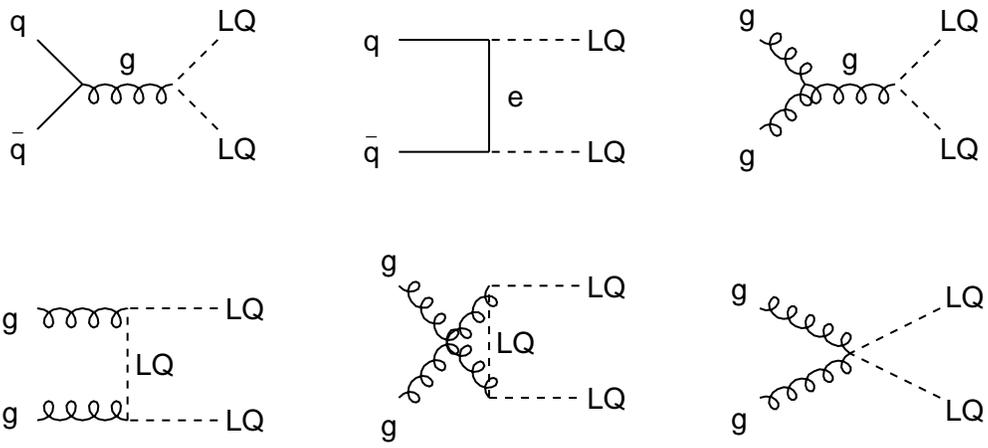
\begin{figure}[tbp]
\caption{Feynman diagrams for leptoquark pair production via $q \bar{q}$
annihilation and gluon fusion.}
\label{feynlq}
\end{figure}

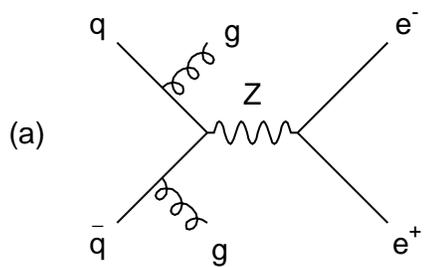
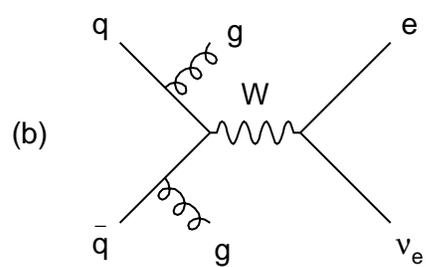
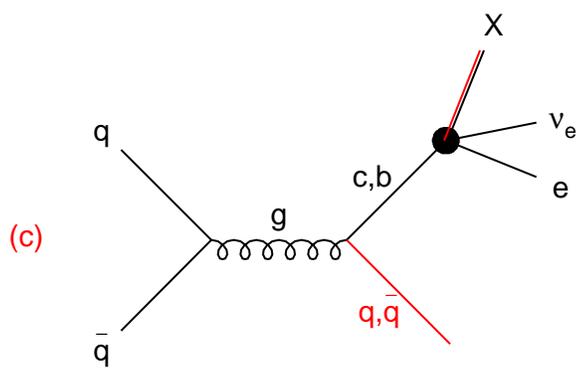
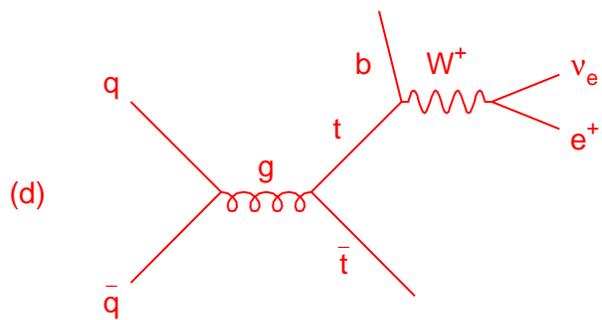
\begin{figure}[tbp]
\caption{Examples of Feynman diagrams for (a) $Z^{*}jj$, (b) $W^{*}jj$, (c)
QCD processes and (d) $t \bar{t}$ production.}
\label{feynbg}
\end{figure}

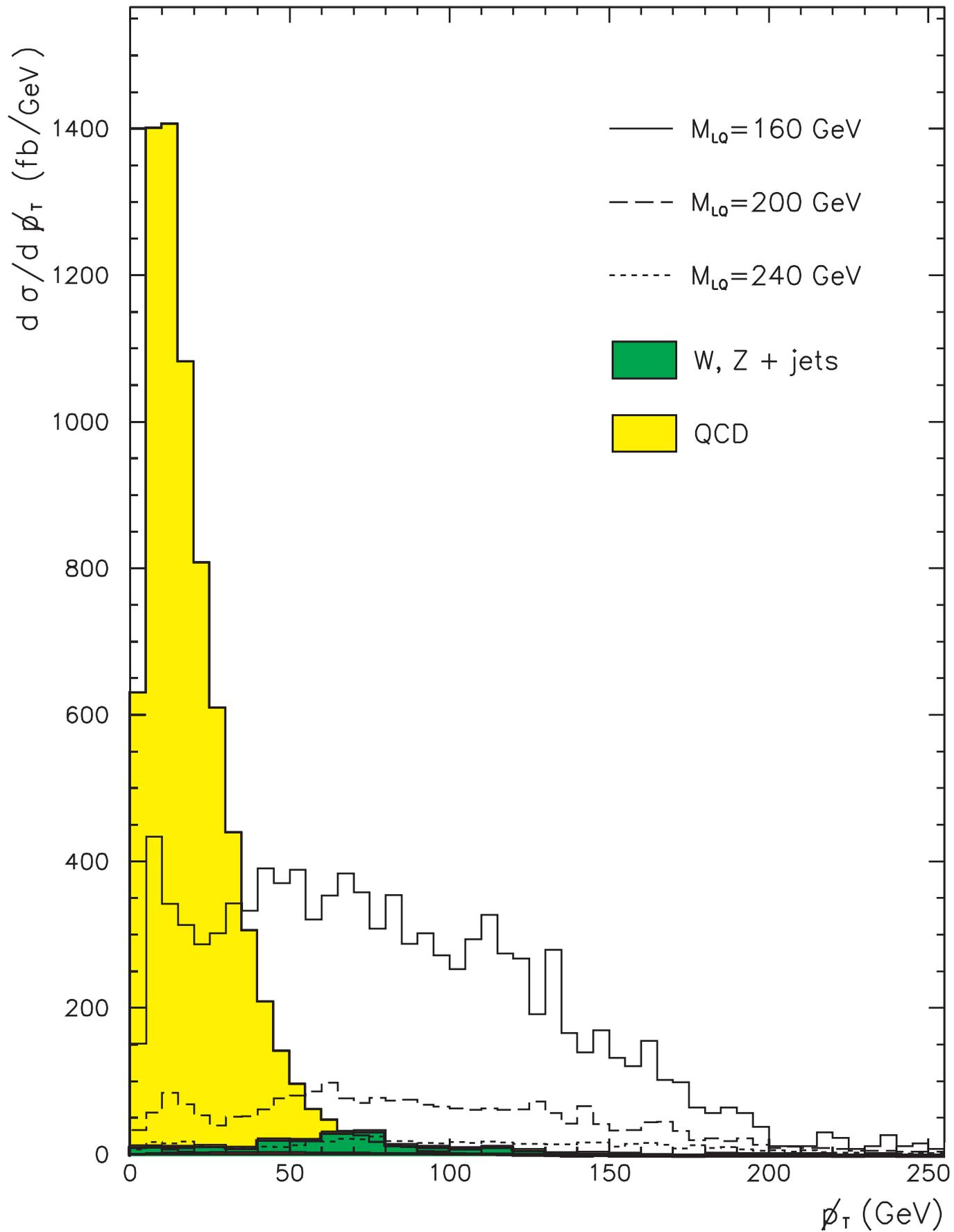
\begin{figure}[tbp]
\caption{Missing $p_T$ (${\not{p}}_T$) distribution in the single electron
channel at the Tevatron, with a branching ratio $B(LQ\rightarrow eq)=0.5$.
The vector leptoquark signal ($\kappa=1$) is displayed for masses $%
M_{LQ}=160 $, 200 and 240 GeV along with the background from $W, Z \, +$
jets and QCD processes. }
\label{pttev}
\end{figure}

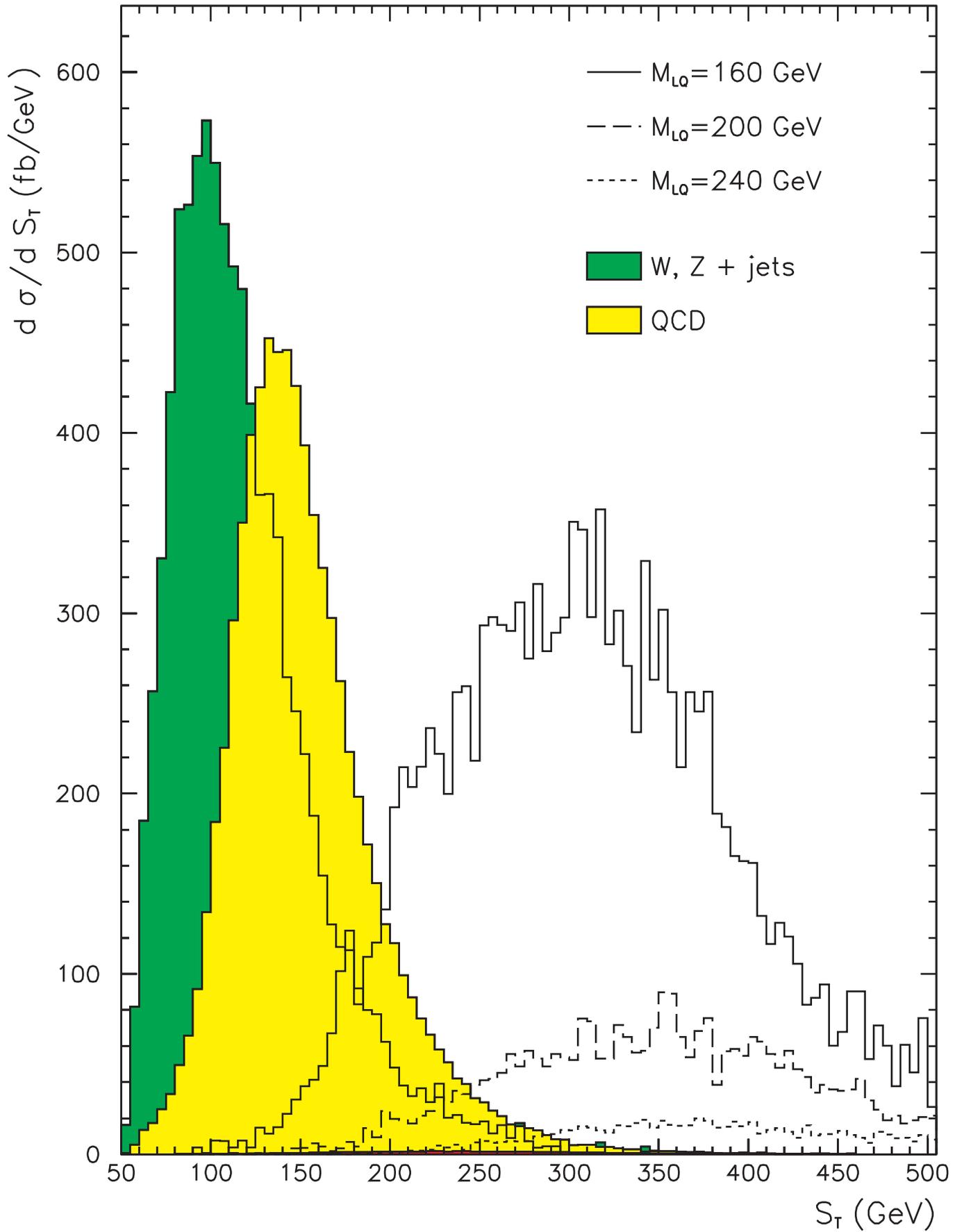
\begin{figure}[tbp]
\caption{Total transverse energy ($S_T$) distribution at the Tevatron, with
a branching ratio $B(LQ\rightarrow eq)=0.5$. Contributions from the
dielectron and single-electron channels have been added and a missing
transverse energy cut of ${\not{p}}_T \geq 40$ GeV has been applied to
events in the single-electron channel. The vector leptoquark signal ($%
\kappa=1$) is displayed for masses $M_{LQ}=160$, 200 and 240 GeV along with
the background from $W, Z \, +$ jets and QCD processes. }
\label{sttev}
\end{figure}

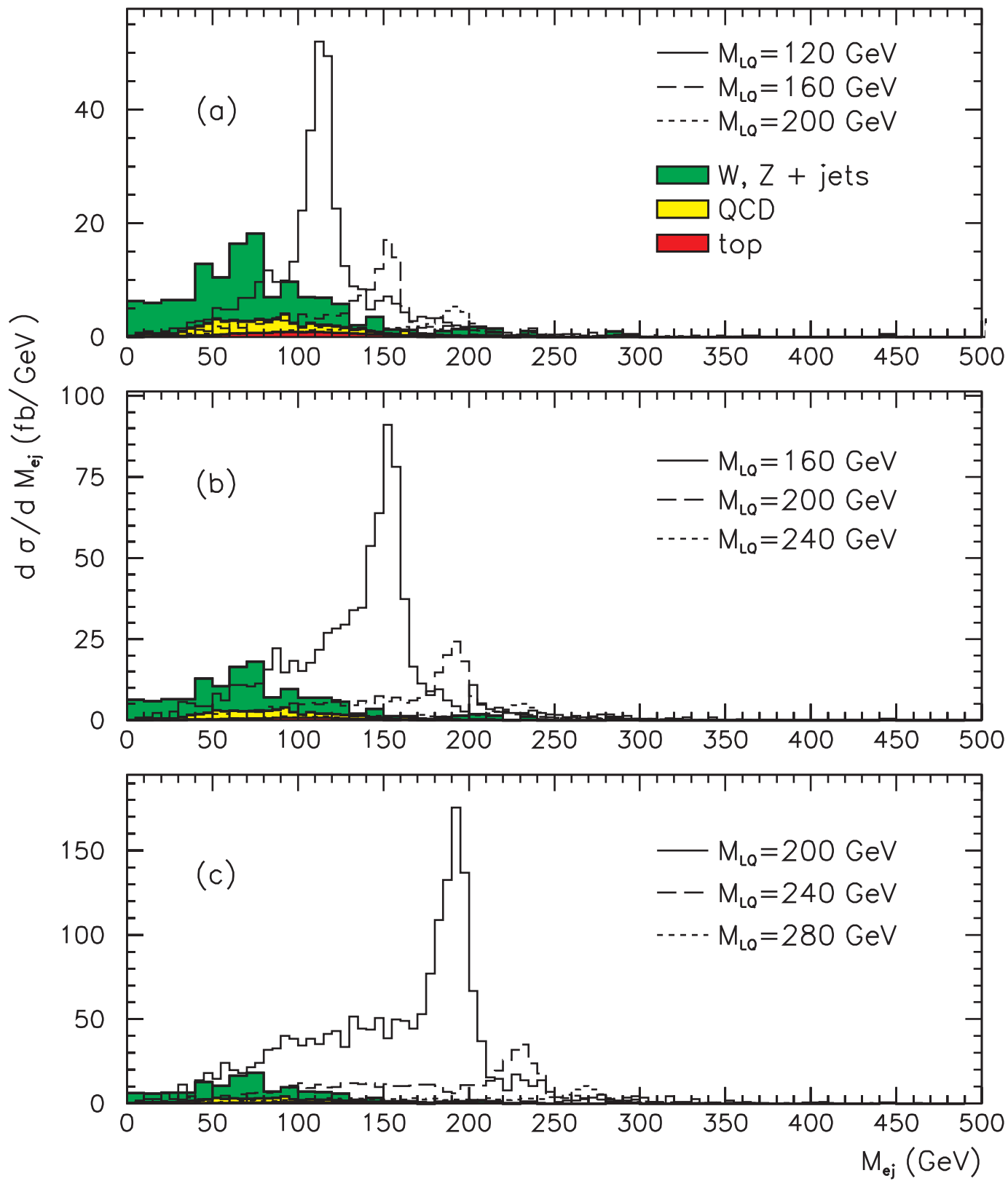
\begin{figure}[tbp]
\caption{Distribution of the invariant mass of the lepton-jet pair for (a)
scalar (b) vector with $\kappa=0$ and (c) vector with $\kappa=1$ leptoquarks
and for the various backgrounds at the Tevatron. Contributions from the
dielectron and single-electron channels have been added and cuts have been
applied on ${\not{p}}_T$ and $S_T$.}
\end{figure}

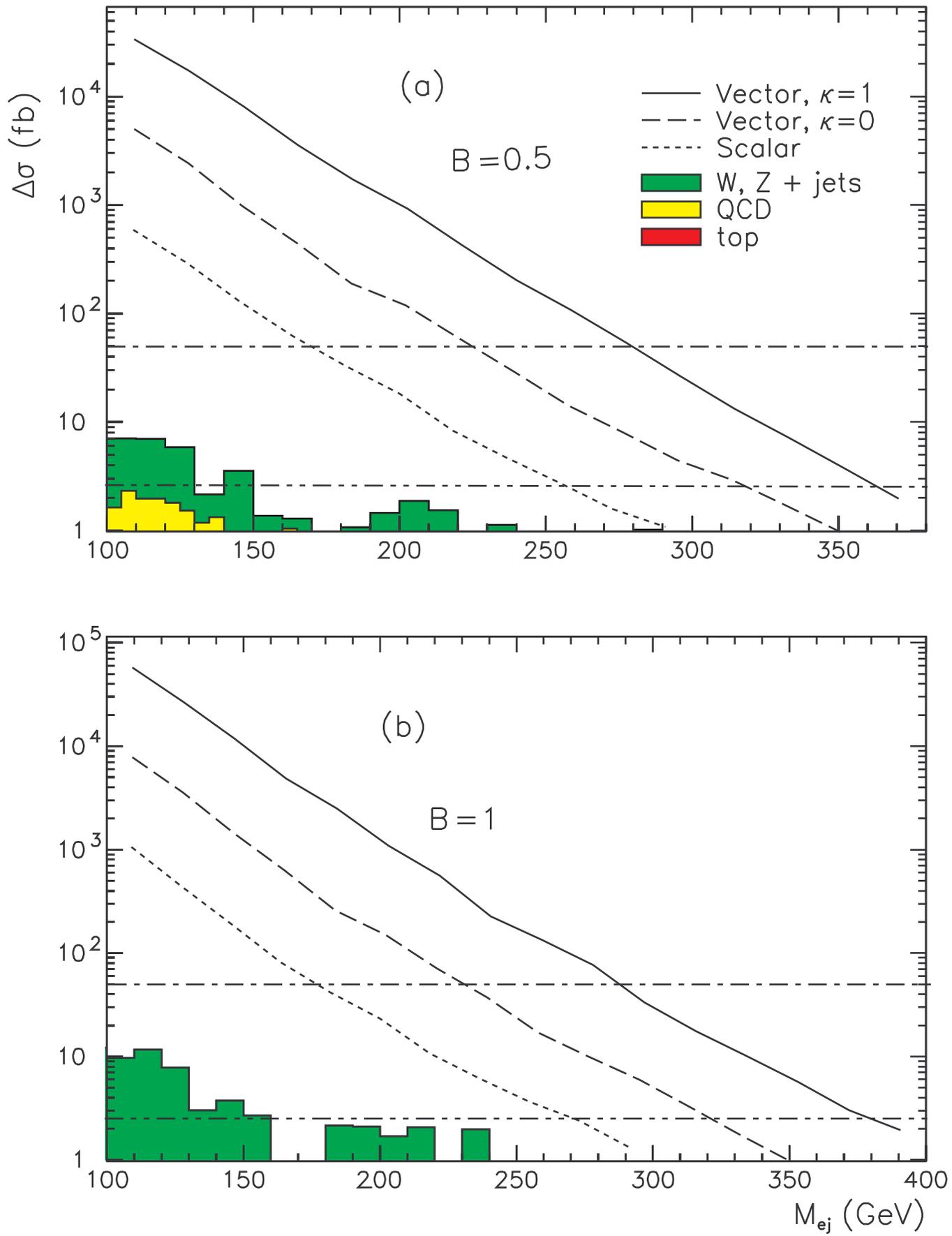
\begin{figure}[tbp]
\caption{Partial cross section integrated around the leptoquark mass peak as
a function of the invariant mass of the electron-jet pair for scalar and
vector ($\kappa=0,1$) leptoquarks and for the various backgrounds at the
Tevatron with (a) $B(LQ\rightarrow eq)=0.5$ and (b) $B(LQ\rightarrow eq)=1$.
Kinematic cuts have been applied on ${\not{p}}_T$ and $S_T$. The dash-dotted
(dash-dot-dotted) lines correspond to the observation of 5 events
considering a
luminosity of 100 pb$^{-1}$ (2 fb$^{-1}$).}
\end{figure}

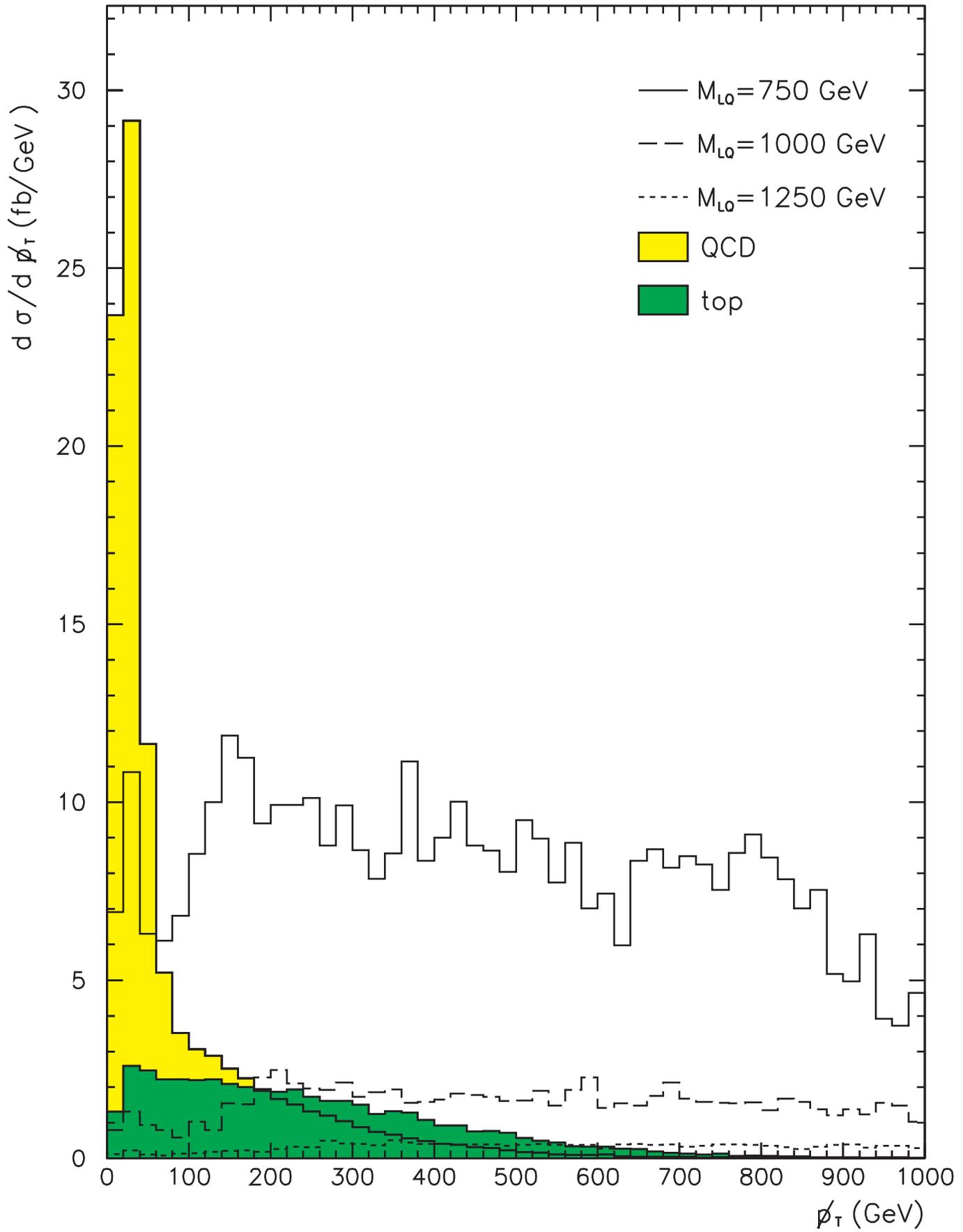
\begin{figure}[tbp]
\caption{Missing $p_T$ (${\not{p}}_T$) distribution in the single electron
channel at the LHC, with a branching ratio $B(LQ\rightarrow eq)=0.5$. The
vector leptoquark signal ($\kappa=1$) is displayed for masses $M_{LQ}=750$,
1000 and 1250 GeV along with the background from QCD processes and $t \bar t$
production. }
\label{ptlhc}
\end{figure}

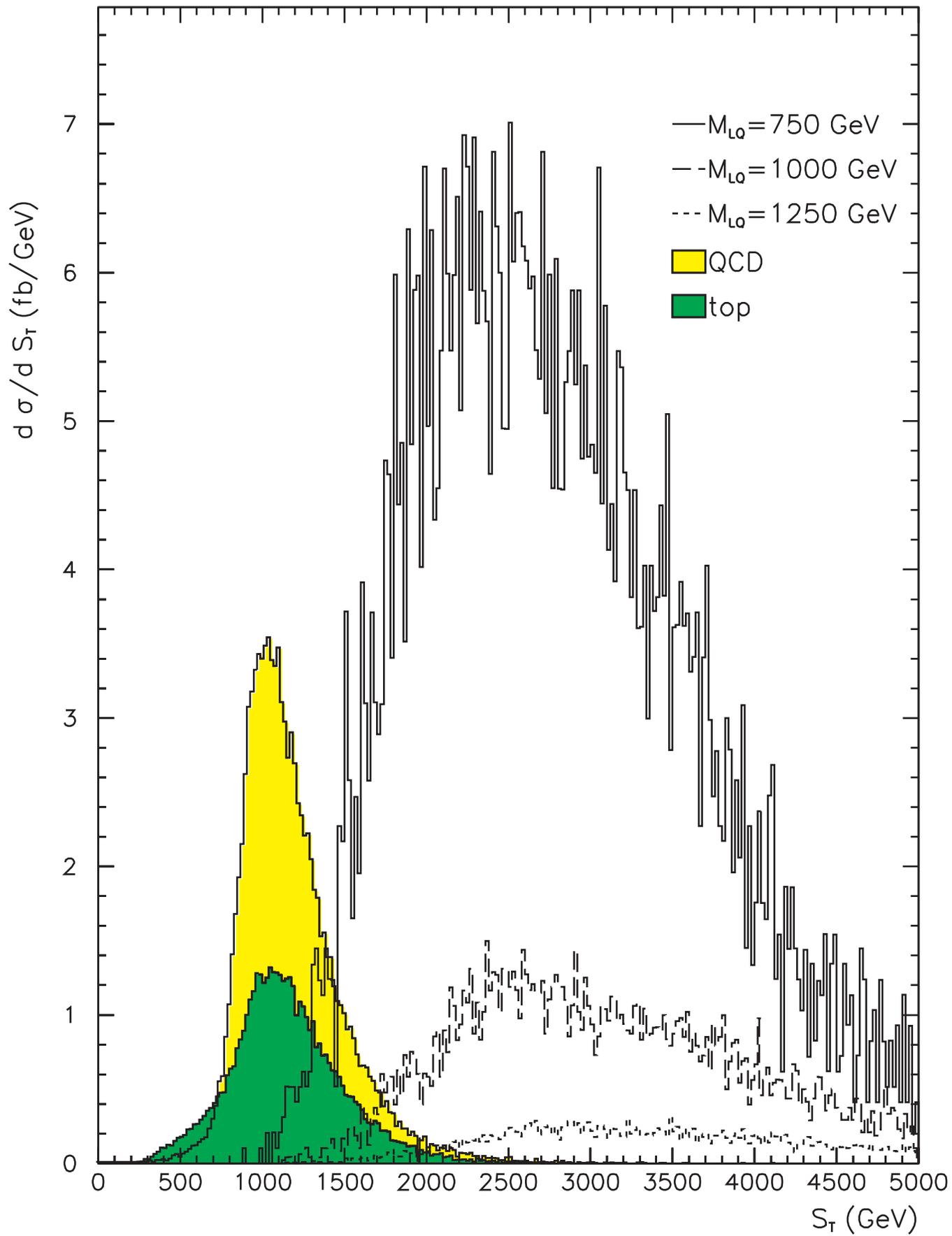
\begin{figure}[tbp]
\caption{Total transverse energy ($S_T$) distribution at the LHC, with a
branching ratio $B(LQ\rightarrow eq)=0.5$. Contributions from the dielectron
and single-electron channels have been added and a missing transverse energy
cut of ${\not{p}}_T \geq 200$ GeV has been applied to events in the
single-electron channel. The vector leptoquark signal ($\kappa=1$) is
displayed for masses $M_{LQ}=750$, 1000 and 1250 GeV along with the
background from QCD processes and $t \bar t$ production. }
\label{stlhc}
\end{figure}

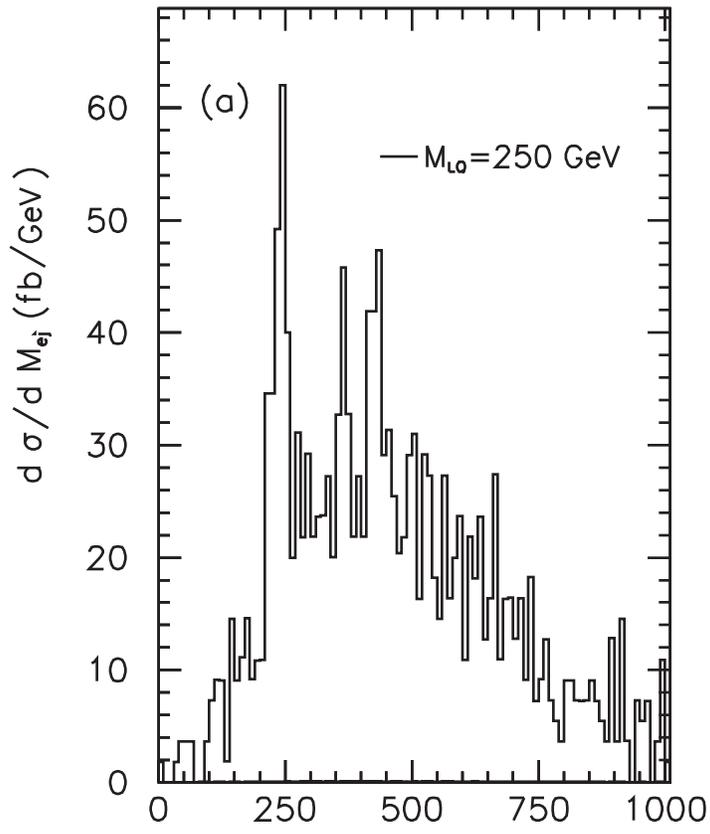
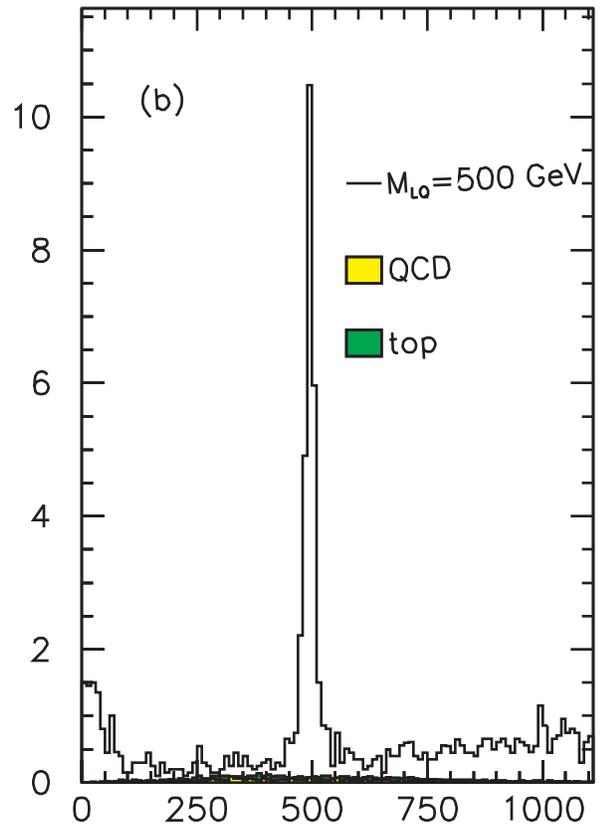
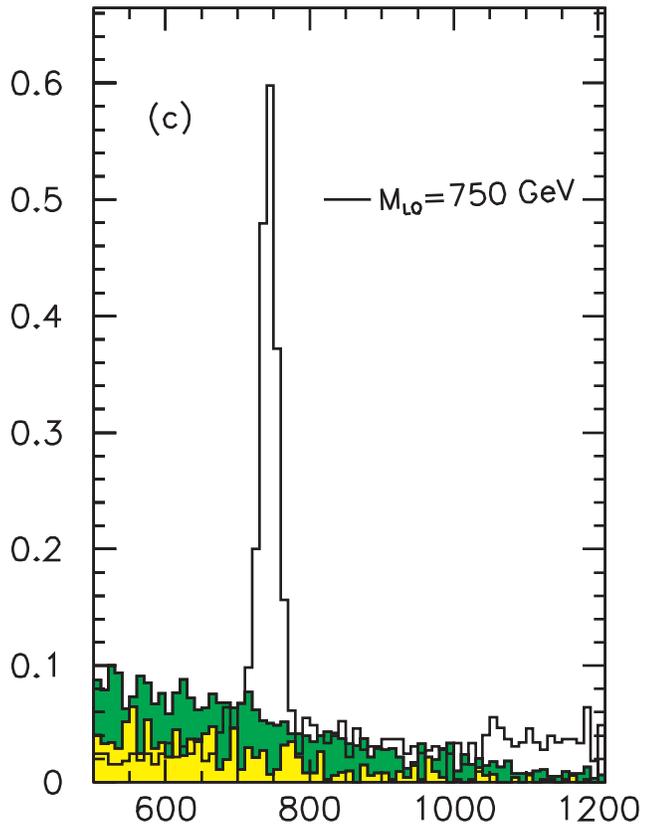
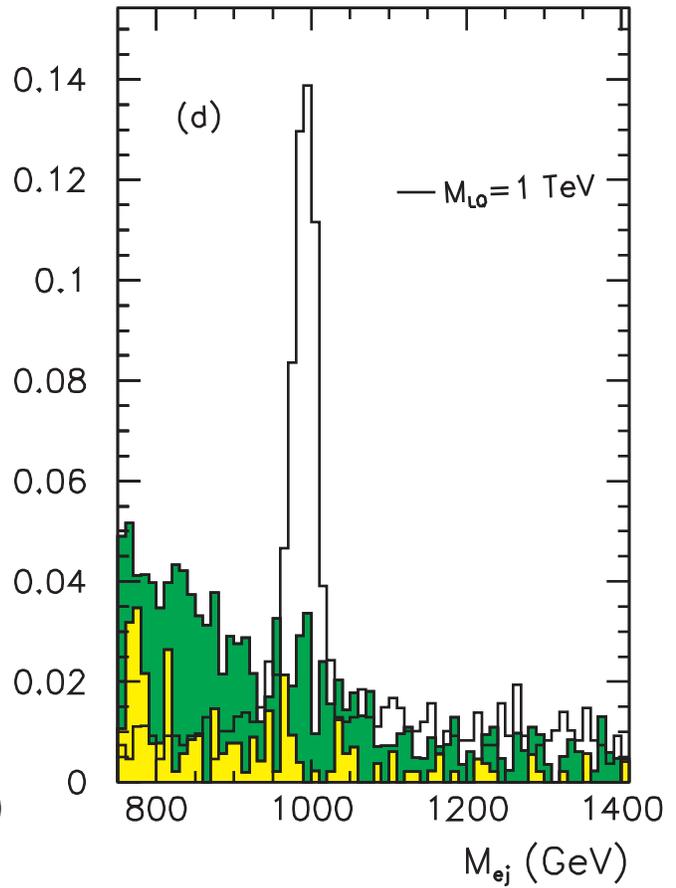
\begin{figure}[tbp]
\caption{Distribution of the invariant mass of the lepton-jet pair for
scalar leptoquarks of mass (a) 250, (b) 500, (c) 750 and (d) 1000 GeV and
for the various backgrounds at the LHC. Contributions from the dielectron
and single-electron channels have been added and cuts have been applied on ${%
\not{p}}_T$ and $S_T$.}
\end{figure}

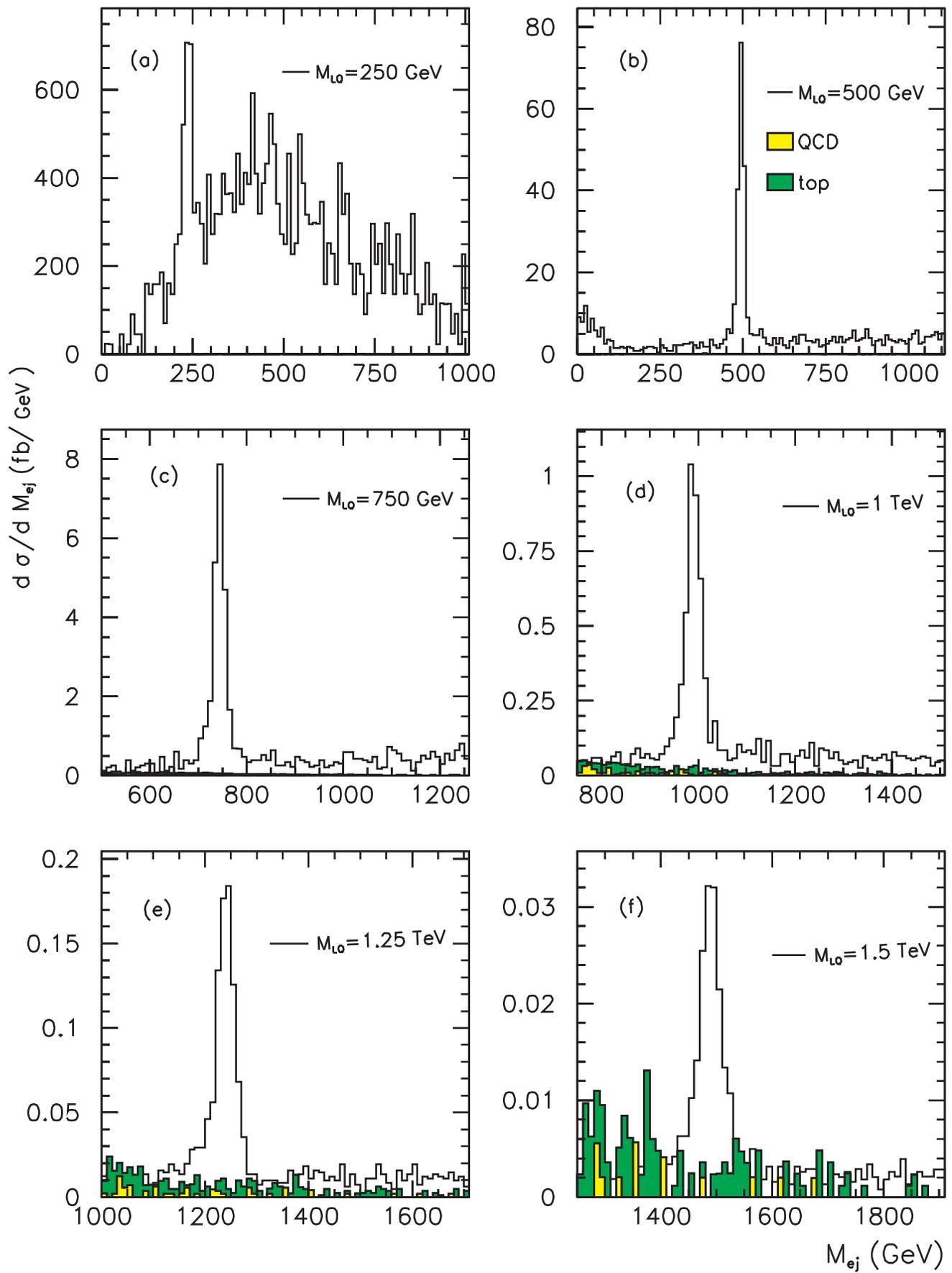
\begin{figure}[tbp]
\caption{Distribution of the invariant mass of the lepton-jet pair for
vector leptoquarks with $\kappa=0$ of mass (a) 250, (b) 500, (c) 750, (d)
1000, (e) 1250 and (f) 1500 GeV and for the various backgrounds at the LHC.
Contributions from the dielectron and single-electron channels have been
added and cuts have been applied on ${\not{p}}_T$ and $S_T$.}
\end{figure}

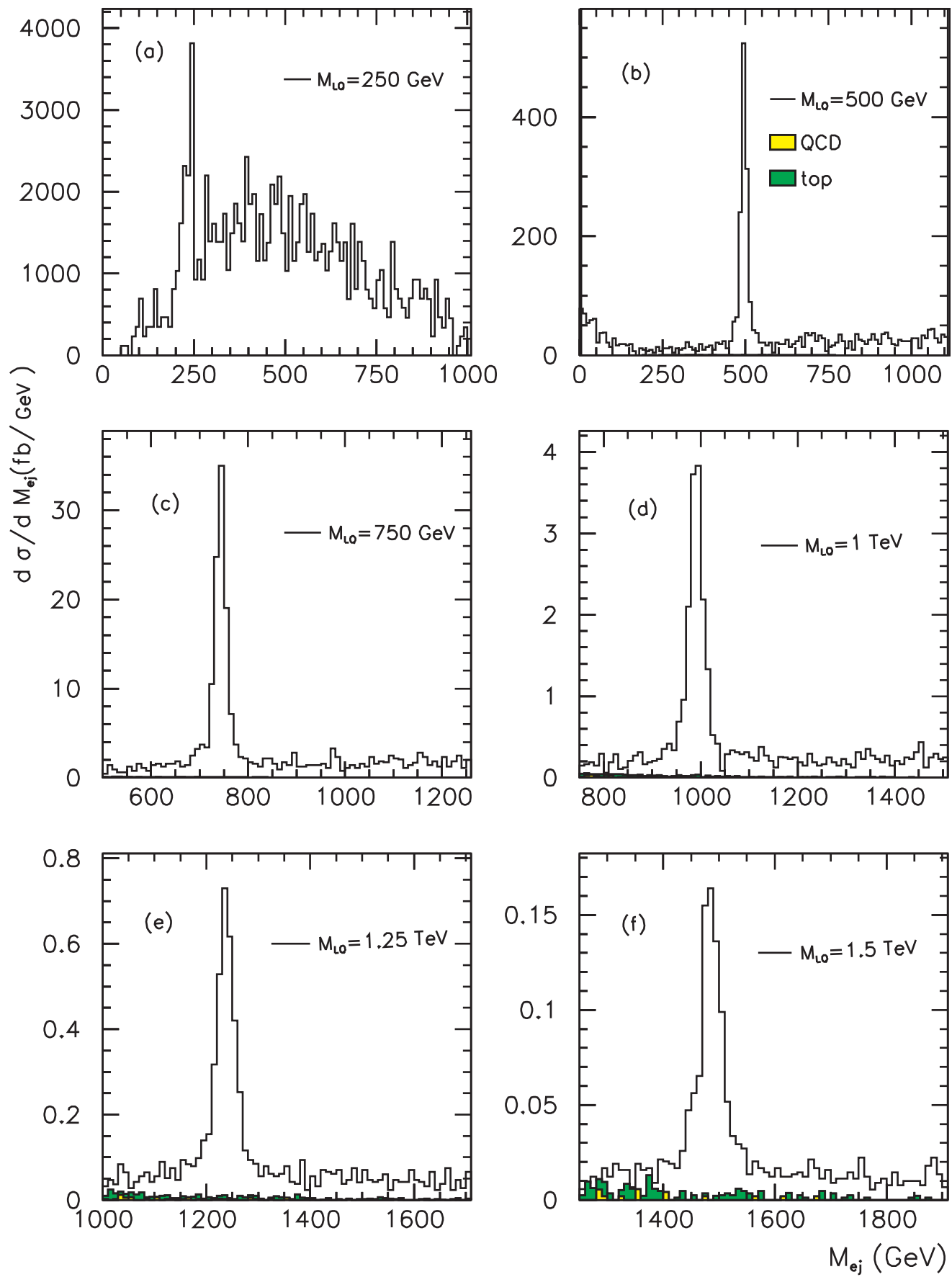
\begin{figure}[tbp]
\caption{Same as Fig. 10 for vector leptoquarks with $\kappa=1$.}
\end{figure}

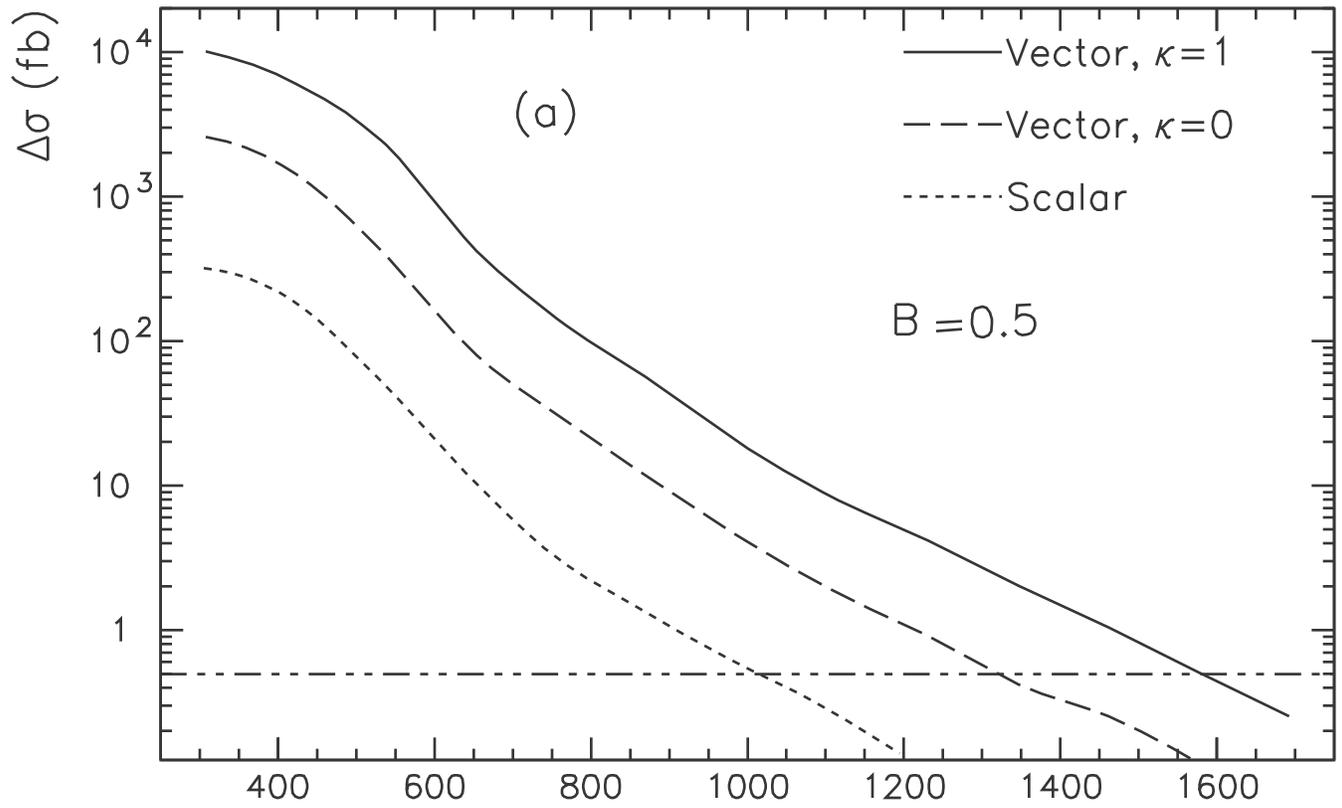
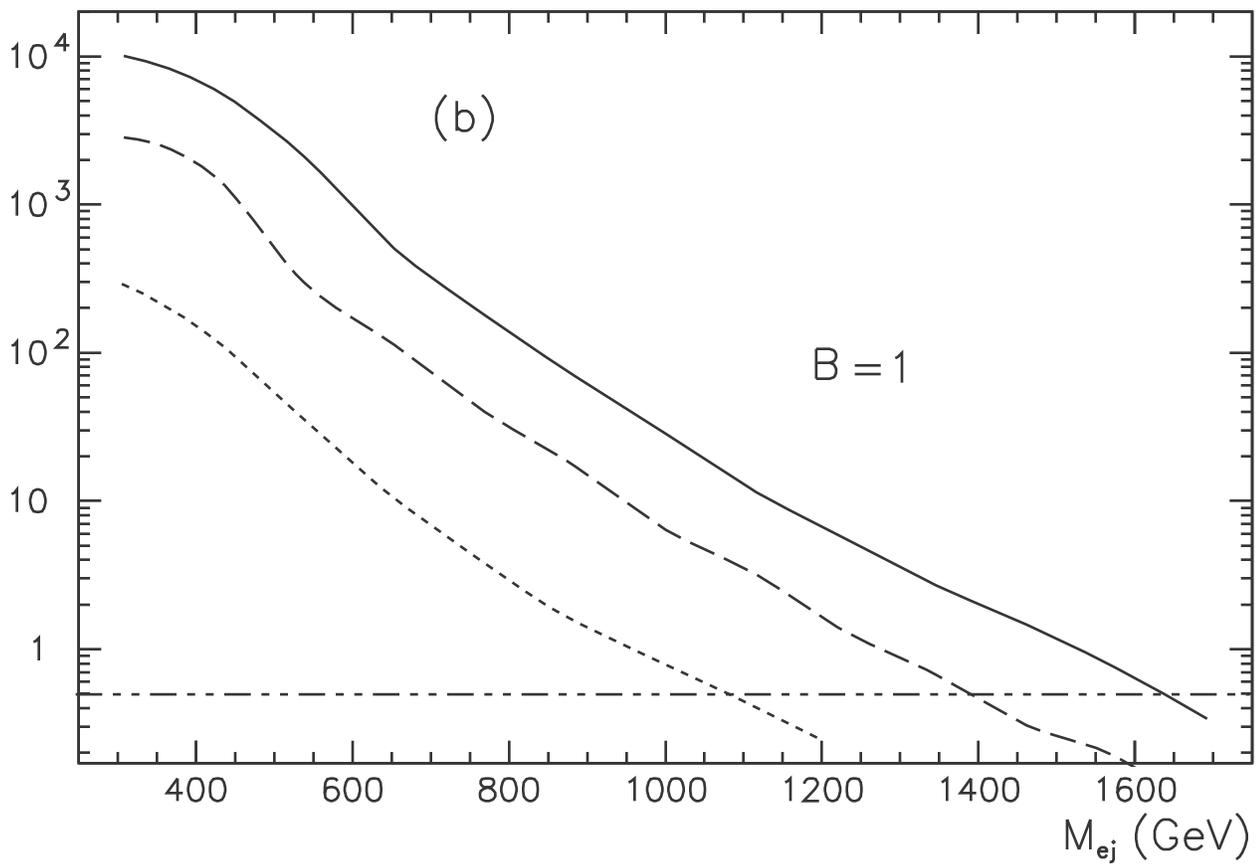
\begin{figure}[tbp]
\caption{Partial cross section integrated around the leptoquark mass peak as
a function of the invariant mass of the electron-jet pair for scalar and
vector ($\kappa=0,1$) leptoquarks and for the various backgrounds at the LHC
with (a) $B(LQ\rightarrow eq)=0.5$ and (b) $B(LQ\rightarrow eq)=1$.
Kinematic cuts have been applied on ${\not{p}}_T$ and $S_T$. The
dash-dot-dotted
lines correspond to the observation of 5 events considering a luminosity of
10 fb$^{-1}$.}
\end{figure}

\end{document}